\documentclass[english]{elsarticle}
\usepackage[T1]{fontenc}
\usepackage[latin9]{inputenc}
\usepackage{amsmath}
\usepackage{amssymb}
\usepackage{babel}
\begin{document}

\title{Electron-positron  pair production in ion collisions at low velocity beyond Born approximation.}
\author{R.N. Lee and A.I. Milstein}
\address{Budker Institute of Nuclear Physics, 630090 Novosibirsk, Russia}
\begin{abstract}
We derive the spectrum and the total cross section of electromagnetic $e^{+}e^{-}$ pair production in the collisions of two nuclei at low relative velocity $\beta$.
Both free-free and bound-free $e^{+}e^{-}$ pair production is considered.
The parameters $\eta_{A,B}=Z_{A,B}\alpha$  are assumed to be small compared to unity but arbitrary compared to $\beta$ ($Z_{A,B}$ are the charge numbers of the nuclei and $\alpha$ is the fine structure constant).
Due to a suppression of the Born term by high power of $\beta$, the first Coulomb correction to the amplitude appears to be important at $\eta_{A,B}\gtrsim \beta$.
The effect of a finite nuclear mass is discussed.
In contrast to the result obtained in the infinite nuclear mass limit, the terms $\propto M^{-2}$ are not suppressed by the high power of $\beta$ and may easily dominate at sufficiently small velocities.
\end{abstract}
\maketitle
\section{Introduction}

The process of electromagnetic $e^{+}e^{-}$ pair production in heavy-ion collisions plays an essential role in collider experiments.
It has a long history of experimental and theoretical investigations.
The process takes place in two different flavors dubbed as ``free-free'' and ``bound-free'' production, depending on whether the final electron is in the continuous spectrum or in the bound state with one of the nuclei.

As for the free-free pair production, the pioneering papers \cite{LandLif1934,Racah1936} appeared already in 1930-s and dealt with the high-energy asymptotics of the process.
In late 1990-s the interest to the process has been revived due to the RHIC experiment and approaching launch of the LHC experiment.
In particular, the contribution of the higher orders in the parameters $\eta_{A,B}=Z_{A,B}\alpha$ (the Coulomb corrections) in the high-energy limit has been discussed intensively, see Refs. \cite{BaltMcL1998,IvaScSe1999,EiReScG1999,SegeWel1999,Lee2000} and the review \cite{BauHeTr2007}.

The interest to the lepton pair production in collisions of slow nuclei appeared long ago in connection with the supercritical regime taking place when the total charge of the nuclei is large enough (at least larger than 173), see Ref. \cite{Greiner1985} and references therein.

Recently, in Ref. \cite{LeeMingulov2016} the total Born cross section of the free-free pair production has been calculated exactly in the relative velocity $\beta$ of the colliding nuclei.
It turns out that the cross section is strongly suppressed as $\beta^8$ at $\beta\ll 1$. A natural question arises whether such a suppression also holds for the Coulomb corrections (the higher terms in $\eta_{A,B}$).

In the present paper we show that the Coulomb corrections are less suppressed with respect to $\beta$ than the Born term.
We assume that $\beta\ll1$ and $\eta_{A,B}\ll1$ and take into account the higher-order terms in $\eta_{A,B}$ amplified with respect to $\beta$.
We consider both free-free and bound-free pair production.
In the next section we perform calculations in the approximation in which both nuclei have constant velocities, i.e., we treat the nuclei as infinitely heavy objects and neglect the Coulomb interaction between them.
This approach has severe restrictions with respect to values of $\beta$.
These restrictions are discussed in the  third section together with the qualitative modification of the results in the region where the constant-velocity approximation is not valid.

\section{Pair production cross section.}

Let us first assume that the parameter $\eta_{A}$ is sufficiently small to be treated in the leading order.
In particular, we  assume that $\eta_{A}\ll\eta_{B},\beta$.
We neglect the Coulomb interaction between the nuclei and  work in the rest frame of the nucleus $B$ with $z$ axis directed along the momentum of the nucleus $A$.
Since our primary goal is the total cross section of the process, we find it convenient to use the eigenfunctions of angular momentum as a basis.
In this basis the cross section has the form
\begin{align}
d\sigma & =\frac{1}{\beta}2\pi\delta\left(\beta q_{z}-\varepsilon-\tilde{\varepsilon}\right)\sum\left|M\right|^{2}\frac{d^{3}\boldsymbol{q}}{\left(2\pi\right)^{3}}\frac{dp}{2\pi}\frac{d\tilde{p}}{2\pi}\label{eq:sigmaFF}
\end{align}
where $\beta$ is the relative velocity of the nuclei,  $\boldsymbol{q}$ is the space components of the momentum transfer to nucleus $A$, $\varepsilon=\sqrt{p^{2}+m^{2}}$ is the electron energy,  $m$ is the electron mass, and the corresponding  quantities with tildes  are  related to a positron.
The summation  in Eq. (\ref{eq:sigmaFF}) is performed over all discrete quantum numbers related to the states of both particles, i.e., over the total angular momentum $J$ , its projection $M$, and two possible values of $L=J\pm1/2$, related to the parity of the state.
Within our accuracy, the matrix element $M$ reads
\begin{equation}
M=\frac{4\pi\eta_{A}}{\omega\boldsymbol{q}^{2}}\int d\boldsymbol{r}e^{i\boldsymbol{q}\cdot\boldsymbol{r}} \boldsymbol{q}\cdot \boldsymbol{J}\,, \quad \boldsymbol{J}=U^+\left(\eta_{B},\kappa,\varepsilon|\boldsymbol{r}\right)\boldsymbol{\alpha}\,V\left(\eta_{B},\tilde{\kappa}, \tilde{\varepsilon}|\boldsymbol{r}\right)\,.\label{eq:matrixElement}
\end{equation}
Here $\omega=\varepsilon+\tilde{\varepsilon}$,  $U\left(\eta_{B},\kappa,\varepsilon|\boldsymbol{r}\right)$
is the electron wave function with the energy $\varepsilon$, the total angular momentum $J=\left|\kappa\right|-\frac{1}{2}$, and $L=J+\frac{1}{2}\mathrm{sgn}\,\kappa$. This wave function is the solution of the Dirac equation in the attractive potential $-\eta_{B}/r$.
The function $V\left(\eta_{B},\tilde\kappa,\tilde\varepsilon|\boldsymbol{r}\right)$
is the negative-energy solution of the Dirac equation corresponding to the charge conjugation of the positron wave function, so that
\[
V\left(\eta_{B},\tilde\kappa,\tilde\varepsilon|\boldsymbol{r}\right)=i\gamma_{2}U^{*}\left(-\eta_{B},\tilde\kappa,\tilde\varepsilon|\boldsymbol{r}\right)\,.
\]
In the derivation of Eq.~\eqref{eq:matrixElement} we have used the gauge in which the photon propagator has the form $$D^{ab}=-{4\pi(\delta^{ab}-q^aq^b/\omega^2)\over \omega^2-q^2}\ , \quad D^{0a}=D^{00}=0\,.$$

The limit  $\beta\ll 1$ is quite special.
From kinematic constraints, it is easy to conclude (cf. Ref. \cite{LeeMingulov2016}) that the characteristic momentum transfers to both nuclei are of the order of $m/\beta\gg m$.
A simple estimate $r\sim\beta/m\ll1/m\eta_{B}$ justifies using the small-$r$ asymptotics of the Coulomb wave functions:
\begin{align}
U\left(\eta_{B},\kappa,\varepsilon|\boldsymbol{r}\right)&=\left(\begin{array}{c}
f\left(r\right)\Omega_{\kappa M}\\
-ig\left(r\right)\left(\boldsymbol{\sigma}\boldsymbol{n}\right)\Omega_{\kappa M}
\end{array}\right)\,,\nonumber\\
V\left(\eta_{B},\tilde\kappa,\tilde\varepsilon|\boldsymbol{r}\right)&=\left(\begin{array}{c}
\tilde{g}\left(r\right)\Omega_{-\tilde{\kappa}M}\\
-i\tilde{f}\left(r\right)\left(\boldsymbol{\sigma}\boldsymbol{n}\right)\Omega_{-\tilde{\kappa}M}
\end{array}\right)\,,
\end{align}
where $\boldsymbol{n}=\boldsymbol{r}/r$, $\Omega_{\kappa M}=\Omega_{JLM}$ is the spherical spinor, and the radial wave functions read
\begin{align}
\left.\begin{array}{c}
f\left(r\right)\\
g\left(r\right)
\end{array}\right\}  & =Cr^{\gamma-1}\left\{ \begin{array}{c}
\frac{\kappa-\gamma}{\eta_{B}}+\underline{\frac{r}{2\gamma+1}\left[\varepsilon\left(2\gamma-2\kappa+1\right)+m\right]}\\
1-\underline{\frac{r\left(\kappa-\gamma\right)}{\left(2\gamma+1\right)\eta_{B}}\left[\varepsilon\left(2\gamma+2\kappa+1\right)-m\right]}
\end{array}\right. \label{eq:radialFunctions}\nonumber\\
\left.\begin{array}{c}
\tilde{f}\left(r\right)\\
\tilde{g}\left(r\right)
\end{array}\right\}  & =\tilde{C}r^{\tilde{\gamma}-1}
\left\{
\begin{array}{c}
-\frac{\tilde{\kappa}-\tilde{\gamma}}{\eta_{B}}+\underline{\frac{r}{2\tilde{\gamma}+1}\left[\tilde{\varepsilon}\left(2\tilde{\gamma}-2\tilde{\kappa}+1\right)+m\right]}\\
1+\underline{\frac{r\left(\tilde{\kappa}-\tilde{\gamma}\right)}{\left(2\tilde{\gamma}+1\right)\eta_{B}}\left[\tilde{\varepsilon}\left(2\tilde{\gamma}+2\tilde{\kappa}+1\right)-m\right]}
\end{array}\right.
\end{align}
Here
\begin{gather*}
C=\frac{p}{\varepsilon}\sqrt{\frac{1+\frac{\gamma}{\kappa}}{1+\frac{m\gamma}{\varepsilon\kappa}}}e^{\pi\nu/2}\left|\frac{\Gamma\left(\gamma+1+
i\nu\right)}{\Gamma\left(2\gamma+1\right)}\right|\left(2p\right)^{\gamma},\\
\gamma=\sqrt{\kappa^{2}-\eta_{B}^{2}}\,,\quad\nu=\varepsilon\eta_{B}/p\,,\\
\tilde{C}=C\left(\eta_{B}\to-\eta_{B},\ \varepsilon\to\tilde{\varepsilon},\ \kappa\to\tilde{\kappa}\right)\,.
\end{gather*}

Let us assume for the moment that $\beta\ll\eta_{B}\lesssim1$.
Then the underlined terms can be safely neglected due to the estimate $r\sim\beta/m$.
Moreover, due to the same estimate, the leading contribution to the sum in Eq. (\ref{eq:sigmaFF}) is given by the terms with $\kappa=\pm1$
and $\tilde{\kappa}=\pm1$.
If we also assume  that $\eta_{B}\ll1$, then only the contributions of two states with
\[
\left(\kappa,\tilde{\kappa}\right)=\left(+1,-1\right)\text{ and }\left(\kappa,\tilde{\kappa}\right)=\left(-1,+1\right)
\]
 survive.
The underlined terms in Eq.~\eqref{eq:radialFunctions} become important for $\eta_{B}\lesssim\beta$.
In this region, in addition to the two   states mentioned above,   the states with $\left(\kappa,\tilde{\kappa}\right)$ equal to
\[
\left(+1,+2\right)\ ,\quad\left(-1,-2\right)\ ,\quad\left(+2,+1\right)\ ,\quad\text{and}\quad\left(-2,-1\right)
\]
also should be taken into account.
Integrating over $\boldsymbol{r}$ in Eq.~\eqref{eq:matrixElement}, substituting the result in Eq.~(\ref{eq:sigmaFF}), and integrating over $\boldsymbol{q}$, we obtain the cross section $\sigma_{ff}$ of the free-free pair production:
\begin{multline}
\frac{d\sigma_{ff}}{d\varepsilon d\tilde{\varepsilon}} =\frac{\eta_{A}^{2}\eta_{B}^{2}\beta^{6}p\tilde{p}}{\pi\left(\tilde{\varepsilon}+\varepsilon\right)^{8}}\Bigg\{\pi^{2}\eta_{B}^{2}\left(\varepsilon\tilde{\varepsilon}-m^{2}\right)
-\frac{128\pi\eta_{B}\beta(\tilde{\varepsilon}-\varepsilon)}{27(\tilde{\varepsilon}+\varepsilon)}\left(\varepsilon\tilde{\varepsilon}-2m^{2}\right)
\\
 +\frac{16\beta^{2}}{45\left(\tilde{\varepsilon}+\varepsilon\right)^{2}}\left[\left(33\varepsilon\tilde{\varepsilon}-49m^{2}
 \right)\left(\varepsilon^{2}+\tilde{\varepsilon}^{2}\right)-14\varepsilon^{2}\tilde{\varepsilon}^{2}+78m^{2}\varepsilon\tilde{\varepsilon}-32m^{4}\right]\Bigg\}\,.\label{eq:ffSpectrum}
\end{multline}

The relative order of the three terms in braces is regulated by the ratio $\eta_{B}/\beta$.
When this ratio is small, the last term dominates.
This term coincides with the Born result obtained in Ref. \cite{LeeMingulov2016}, as should be.
The parameter $\eta_{B}/\beta$ appears  due to the ``accidental''
suppression of the Born amplitude of pair production and has nothing to do with  the Sommerfeld-Gamov-Sakharov factor.

The bound-free pair production can be treated exactly in the same way as the free-free pair production.
It appears that an electron is produced mostly in $ns_{1/2}$
states ($\kappa=-1$).
The positron spectrum reads
\begin{multline}
\frac{d\sigma_{bf}}{d\tilde{\varepsilon}}  =\eta_{A}^{2}\eta_{B}^{5}\beta^{6}\frac{2m^{3}\left(\tilde{\varepsilon}-m\right)\tilde{p}}{\left(\tilde{\varepsilon}+m\right)^{8}}\zeta_{3}\Bigg\{\pi^{2}\eta_{B}^{2}-
\frac{128\pi\eta_{B}\beta(\tilde\varepsilon-2m)}{27(\tilde{\varepsilon}+m)} \\
  +\frac{16\beta^{2}}{15\left(\tilde{\varepsilon}+m\right)^{2}}\left[11\tilde{\varepsilon}^{2}-10m\tilde{\varepsilon}+27m^{2}\right]\Bigg\}\,,\label{eq:bfSpectrum}
\end{multline}
where the Riemann zeta function $\zeta_{3}=\sum_{n=1}^{\infty}\frac{1}{n^{3}}$
comes from  summation over the principal quantum number.
It is quite remarkable that Eq.~(\ref{eq:bfSpectrum}) can be obtained from Eq.~(\ref{eq:ffSpectrum})
by the simple substitution  $\frac{p\varepsilon d\varepsilon}{2\pi^{2}}\to\sum_{n}\left|\psi_{ns}\left(0\right)\right|^{2}=\sum_{n}\frac{m^{3}\eta_{B}^{3}}{\pi n^{3}}$
followed by the replacement $\varepsilon\to m$.
This substitution works because of the factorization of  hard-scale $r\sim\beta/m$ and soft-scale $r\sim1/m\eta_{B}$  contributions.

The total cross sections are obtained by the direct integration over energies (energy)\footnote{Note that Eq. \eqref{eq:totalCrossSectionB} is in obvious contradiction with the results of Refs. \cite{Khriplovich2014,Khriplovich2016}. The origin of discrepancy is different for these two papers. As it concerns free-free pair production, in Ref. \cite{Khriplovich2014}  two definitions for the total momentum transfer from the nuclei (differing by the relative sign between momentum transfers from each nucleus) appear to be mixed. Meanwhile, in Ref. \cite{Khriplovich2016} the space components of momentum transfer from the projectile nucleus (of the order of $m/\beta\gg m$!) are totally omitted in the annihilation current.}:
\begin{align}
\sigma_{ff} & =\frac{\eta_{A}^{2}\eta_{B}^{2}\beta^{6}}{1050\pi m^{2}}\Bigg\{\pi^{2}\eta_{B}^{2}+\frac{592}{105}\beta^{2}\Bigg\}\,,\nonumber\\
\sigma_{bf} & =\frac{16\eta_{A}^{2}\eta_{B}^{5}\beta^{6}}{15015m^{2}}\zeta_{3}\left\{ \pi^{2}\eta_{B}^{2}+\frac{976}{153}\beta^{2}\right\} \,.\label{eq:totalCrossSectionB}
\end{align}
The main contribution to the integral is given by the region $p\sim m,\,\tilde p\sim m$.
Note the cancelation of the terms $\propto\eta_{B}$  in braces for both free-free and bound-free cross sections.
While this cancellation for the free-free case is a trivial consequence of the charge parity conservation, for the bound-free case it comes as a sort of surprise.

As it concerns the free-free pair production, the results \eqref{eq:ffSpectrum} and \eqref{eq:totalCrossSectionB} can be reproduced in a completely independent way.
Namely, one can obtain the matrix element of the process in conventional diagrammatic technique taking into account the diagrams shown in Fig. 1. and calculating the contribution of the region where all Coulomb exchanges have momenta $\sim m/\beta$.
\begin{figure}
\centering
\includegraphics[width=0.7\linewidth]{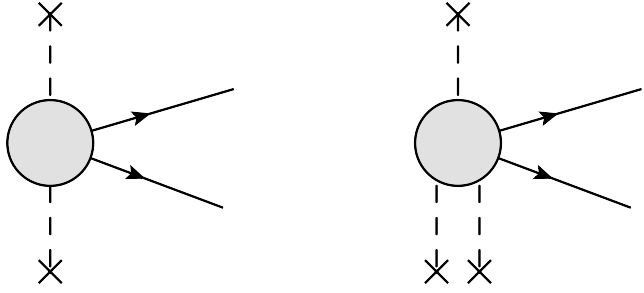}
\caption{Diagrams of the $e^+e^-$ pair production with the account of the first Coulomb correction in $\eta_B$.}
\label{fig:diagramsPP}
\end{figure}

This contribution gives the correct amplitude up to the Coulomb phase which cancels in the cross sections \eqref{eq:ffSpectrum} and \eqref{eq:totalCrossSectionB}.

Let us now assume that $\eta_{A}\sim\eta_{B}$.
Then the higher-order terms in $\eta_{A}$ should be treated on the same basis as those in $\eta_{B}$.
However, the account of these terms is not reduced to the substitution $\eta_{A}\leftrightarrow\eta_{B}$
in Eqs.(\ref{eq:ffSpectrum}), (\ref{eq:bfSpectrum}), and (\ref{eq:totalCrossSectionB}).
Speaking of the free-free pair production, the substitution $\eta_{A}\leftrightarrow\eta_{B}$ should be taken into account on the level of matrix element, but not in the cross section.
The relative phase between the contributions $\propto \eta_A\eta_B^2$ and  $\propto \eta_A^2\eta_B$ to the matrix element can be fixed from the diagrammatic approach mentioned above.
Then we obtain
\begin{align}
\frac{d\sigma_{ff}}{d\varepsilon d\tilde{\varepsilon}} & =\frac{\eta_{A}^{2}\eta_{B}^{2}\beta^{6}p\tilde{p}}{\pi\left(\tilde{\varepsilon}+\varepsilon\right)^{8}}
\Bigg\{\pi^{2}\left(\eta_{A}+\eta_{B}\right)^{2}\left(\varepsilon\tilde{\varepsilon}-m^{2}\right)\nonumber\\
&-
\frac{128\pi\left(\eta_{A}+\eta_{B}\right)\beta(\tilde{\varepsilon}-\varepsilon)}{27(\tilde{\varepsilon}+\varepsilon)}\left(\varepsilon\tilde{\varepsilon}-2m^{2}\right)\nonumber \label{eq:ffSpectrumAndCrossSection}\\
 &+\frac{16\beta^{2}}{45\left(\tilde{\varepsilon}+\varepsilon\right)^{2}}\big[\left(33\varepsilon\tilde{\varepsilon}-49m^{2}\right)\left(\varepsilon^{2}+\tilde{\varepsilon}^{2}\right)-14\varepsilon^{2}\tilde{\varepsilon}^{2}+78m^{2}\varepsilon\tilde{\varepsilon}-32m^{4}\big]\Bigg\}\,,\nonumber \\
\sigma_{ff} & =\frac{\eta_{A}^{2}\eta_{B}^{2}\beta^{6}}{1050\pi m^{2}}\Bigg\{\pi^{2}\left(\eta_{A}+\eta_{B}\right)^{2}+\frac{592}{105}\beta^{2}\Bigg\}\,.
\end{align}

For the bound-free pair production we have
\begin{align}\label{eq:bfSpectrumAndCrossSection}
\frac{d\sigma_{bf}}{d\tilde{\varepsilon}} & =\left(\eta_{A}^{2}\eta_{B}^{5}+\eta_{B}^{2}\eta_{A}^{5}\right)\beta^{6}\frac{2m^{3}\left(\tilde{\varepsilon}-m\right)\tilde{p}}
{\left(\tilde{\varepsilon}+m\right)^{8}}\zeta_{3}\Bigg\{\pi^{2}\left(\eta_{A}+\eta_{B}\right)^{2}\nonumber\\
 & -\frac{128\pi\left(\eta_{A}+\eta_{B}\right)\beta(\tilde{\varepsilon}-2m)}{27(\tilde{\varepsilon}+m)}+
 \frac{16\beta^{2}}{15\left(\tilde{\varepsilon}+m\right)^{2}}\left[11\tilde{\varepsilon}^{2}-10m\tilde{\varepsilon}+27m^{2}\right]\Bigg\}\,,\nonumber\\
\sigma_{bf} & =\frac{16\left(\eta_{A}^{2}\eta_{B}^{5}+\eta_{B}^{2}\eta_{A}^{5}\right)\beta^{6}}{15015m^{2}}\zeta_{3}\left\{ \pi^{2}\left(\eta_{A}+\eta_{B}\right)^{2}+\frac{976}{153}\beta^{2}\right\} \,.
\end{align}
Note that  $\sigma_{ff}\gg\sigma_{bf}$ for $\eta_A,\,\eta_B\ll1$, in contrast to the statement in Ref.\cite{Khriplovich2016}.
It is interesting that in the supercritical case, $Z_A+Z_B>173$, the relation between  $\sigma_{ff}$ and $\sigma_{bf}$ is opposite (see e.g., Ref.\cite{Shabaev2014}), since in this case at $\beta\rightarrow 0$ due to the energy conservation law  the electron can be produced in the bound state but not in the free state.

\section{Account for the finite nuclear mass}

The results \eqref{eq:ffSpectrumAndCrossSection} and \eqref{eq:bfSpectrumAndCrossSection} are obtained in the limit $M_A,\,M_B\rightarrow \infty$.
We show  in this section that the  account for  the finite nuclear mass  leads to an essential modification of both   free-free and bound-free pair production  cross sections at sufficiently small $\beta$.

One of the sources, which restrict the applicability of Eqs.~\eqref{eq:ffSpectrumAndCrossSection} and \eqref{eq:bfSpectrumAndCrossSection}, is a deviation of the nuclear trajectories from the straight lines due to the Coulomb interaction between the nuclei.
This deviation can be neglected if a shift of the minimal distance between the nuclei is smaller than the  impact parameter $\rho$ :
\begin{align}
\frac{Z_AZ_B\alpha}{M_r\beta^2}\ll\rho \,,\quad M_r=\frac{M_AM_B}{M_A+M_B}\,,
\end{align}
where $M_A$ and $M_B$ are the masses of the corresponding nuclei.
Substituting $\rho\sim \beta/m$ we come to the constraint
\begin{align}\label{eq:constr}
\beta \gg \left(\frac{mZ_AZ_B\alpha}{M_r}\right)^{1/3}\sim  \left(\frac{m\eta_{max}}{M_p}\right)^{1/3}   \,,
\end{align}
where $M_p$ is the proton mass and $\eta_{max}=\max\{\eta_A,\,\eta_B\}$.
Let us discuss qualitatively the modification of the cross section at $\beta \lesssim  \left(\frac{m\eta_{max}}{M_p}\right)^{1/3}$.
First of all let us consider the dependence of the cross section $d\sigma_{ff}$ on the impact parameter $\rho$ at  $\beta\ll \eta_{max}\ll 1 $.
Similarly  to the derivation of Eq.~\eqref{eq:ffSpectrumAndCrossSection}, we obtain \begin{align}
\frac{d\sigma_{ff}}{d\varepsilon d\tilde{\varepsilon}d\boldsymbol{\rho}} & =\frac{4\eta_{A}^{2}\eta_{B}^{2}\beta^{4}p\tilde{p}}{9\left(\tilde{\varepsilon}+\varepsilon\right)^{6}}
\left(\eta_{A}+\eta_{B}\right)^{2}(\varepsilon\tilde{\varepsilon}-m^{2})(1+a)^2\,e^{-2a}\,,\nonumber\\
a&=(\tilde{\varepsilon}+\varepsilon)\frac{\rho}{\beta}\,.
\label{eq:ffrho}
\end{align}
Integrating over $\boldsymbol{\rho}$ we obtain the first term in Eq.~\eqref{eq:ffSpectrumAndCrossSection}.
If we consider the classical motion of the nuclei interacting by the Coulomb field, we find that the minimal distance $\rho$  between the nuclei and the relative velocity $\beta$ at this point   are  expressed via the impact parameter $\rho_0$ and the relative velocity $\beta_0$ at infinity as
\begin{align}
\rho=(\sqrt{1+\varkappa^2}+\varkappa)\rho_0\,,\quad \beta=(\sqrt{1+\varkappa^2}-\varkappa)\beta_0\,,\quad \varkappa=\frac{Z_AZ_B\alpha}{M_r\beta^2\rho_0}\,.
\end{align}
Then the parameter $a$ in Eq.~\eqref{eq:ffrho} can be written as \begin{align}
a&=\frac{(\tilde{\varepsilon}+\varepsilon)Z_AZ_B\alpha}{M_r\beta_0^3\,\varkappa}(\sqrt{1+\varkappa^2}+\varkappa)^2\,.
\end{align}
Calculating the minimum value of the quantity $a$ with respect to $\varkappa$, we find that
\begin{align}
a&\geq\frac{3^{3/2}(\tilde{\varepsilon}+\varepsilon)Z_AZ_B\alpha}{M_r\beta_0^3}\,.
\end{align}
Therefore, it follows from Eq.\eqref{eq:ffrho}  that the cross section $\sigma_{ff}$ is exponentially small at $mZ_AZ_B\alpha/(M_r\beta_0^3)\gg 1$.
The same conclusion is valid for the bound-free cross section.

Suppose now that the condition \eqref{eq:constr} holds.
Since the results \eqref{eq:ffSpectrumAndCrossSection} and \eqref{eq:bfSpectrumAndCrossSection} are strongly suppressed by the factor $\beta^6$, it is natural to ask whether  the contributions formally suppressed with respect to $m/M_r$ may dominate at sufficiently small $\beta$.
The answer is positive.
Let us consider the ``bremsstrahlung'' mechanism of pair production, when the pair is produced by a virtual  photon emitted by the scattered nucleus.
We consider the case $Z_AZ_B\alpha/\beta\gg1$ when the motion of the nuclei is classical.
Then the cross section of the $e^+e^-$ pair production can be written as a product of the cross section $\sigma_\gamma$ of bremsstrahlung of virtual photon with the energy $\omega=\tilde{\varepsilon}+\varepsilon$  and the probability of virtual photon conversion into $e^+e^-$ pair.
We assume that $2m<\omega\ll M_r\beta^2$ so that $\sigma_\gamma$ can be calculated in the non-relativistic dipole approximation.
We have \cite{BLP}
\begin{align}
d\sigma_{ff}^{BS}& =\frac{\alpha}{2\pi}\Phi\left(\frac{2m}{\omega}\right)d\sigma_{\gamma}\ \,,\nonumber\\
d\sigma_{\gamma}&=\frac{16\alpha(Z_AZ_B\alpha)^2}{3\,\beta^2}\left(\frac{Z_A}{M_A}-\frac{Z_B}{M_B}\right)^2\,G\left(\frac{\omega}{\omega_0}\right)
\frac{d\omega}{\omega}\,,\nonumber\\
\Phi(x)&=\int\limits_0^{\sqrt{1-x^2}}dt\,\frac{t^2}{1-t^2} \sqrt{1-\frac{x^2}{1-t^2}}\left(1+\frac{x^2}{1-t^2}\right)\left(1-\frac{t^2}{3}\right)\,,
\label{eq:ffbrem}
\end{align}
where $\omega_0= \frac{M_r\beta^3}{Z_AZ_B\alpha}$ and the functions $G(\nu)$ has the following asymptotic forms
\begin{align}
G(x)=\ln(1/x)\,\quad\mbox{for}\quad x\ll 1\,,\nonumber\\
G(x)=\frac{\pi}{\sqrt{3}}\exp(-2\pi x)\,\quad\mbox{for}\quad x\gg 1\,.
\end{align}

It is seen from Eq.~\eqref{eq:ffbrem} that $\sigma_{ff}^{BS}$ is exponentially small for $m\gg\omega_0$, which is in agreement with our previous statement.
For $ m\ll \omega_0$, the main contribution to the integral over $\omega$ is given by the region $m\ll \omega\ll \omega_0$.
Then, taking into account that $\Phi(x)\approx-\frac{2}{3}\ln x$ at $x\ll1$, we obtain  in the leading logarithmic approximation
\begin{align}
\sigma_{ff}^{BS}=\frac{8\eta_A^2\eta_B^2}{27\pi\,\beta^2}\left(\frac{Z_A}{M_A}-\frac{Z_B}{M_B}\right)^2\ln^3\left(\frac{\omega_0}{ m }\right)\,.
\label{eq:ffbremtot}
\end{align}
It is seen that the contribution \eqref{eq:ffbremtot} to $\sigma_{ff}$ starts to dominate over \eqref{eq:ffSpectrumAndCrossSection} very soon as $\beta$ decreases.

\section{Discussion and conclusion}

Let us discuss our results.
We have calculated the infinite-mass limit of the free-free and bound-free pair production cross sections, Eqs.  \eqref{eq:ffSpectrumAndCrossSection} and \eqref{eq:bfSpectrumAndCrossSection}.
As expected, the bound-free pair production cross section is much smaller than the free-free one, with the relative magnitude $\sim \eta^3$.
In the region $\beta\lesssim \eta_{A,B}$  both cross sections essentially deviate from the results obtained in the leading order in $\eta_{A,B}$.
This is due to the ``accidental'' suppression of the Born amplitude.
In this connection, it is interesting to compare Eqs. \eqref{eq:ffSpectrumAndCrossSection} and \eqref{eq:bfSpectrumAndCrossSection} with the corresponding cross sections $\sigma_{ff}^{(0)}$ and $\sigma_{bf}^{(0)}$ for the production of scalar particles.
Using the same technique we easily obtain
\begin{align}\label{eq:scalarffSpectrum}
\frac{d\sigma_{ff}^{(0)}}{d\varepsilon d\tilde{\varepsilon}} & =\frac{16\eta_{A}^{2}\eta_{B}^{2}\beta^{4}p\tilde{p}}{3\pi\left(\tilde{\varepsilon}+\varepsilon\right)^{6}}\,,\quad
\sigma_{ff}^{(0)} =\frac{4\eta_{A}^{2}\eta_{B}^{2}\beta^{4}}{135\pi m^2}\,,\nonumber\\
\frac{d\sigma_{bf}^{(0)}}{d\tilde{\varepsilon}} & =\frac{32\zeta_{3}\eta_{A}^{2}\eta_{B}^{2}(\eta_{A}^{3}+\eta_{B}^{3})\beta^{4}m^{2}\tilde{p}}{3\left(\tilde{\varepsilon}+m\right)^{6}}\,,\quad
\sigma_{bf}^{(0)} =\frac{64\zeta_{3}\eta_{A}^{2}\eta_{B}^{2}(\eta_{A}^{3}+\eta_{B}^{3})\beta^{4}}{315m^2}\,.
\end{align}
The result for $\sigma_{ff}^{(0)}$ coincides with the asymptotics  in Eq. (17) of Ref. \cite{LeeMingulov2016}\footnote{There is an obvious typo in Eq. (17) of Ref. \cite{LeeMingulov2016} --- an extra $\pi$ factor in the denominator.}.
In contrast to the spinor case, the cross sections \eqref{eq:scalarffSpectrum} do not contain the terms of relative order $\eta_{A,B}/\beta$ since the leading-order contribution in $\eta_{A,B}$ is not suppressed by the power of $\beta$ anymore.
We note that the Coulomb corrections in Eqs. \eqref{eq:ffSpectrumAndCrossSection} and \eqref{eq:bfSpectrumAndCrossSection} are still more strongly suppressed in $\beta$ than the Coulomb corrections to the corresponding cross sections for scalar particles, though the suppression is only $\beta^2$, which is to be compared with $\beta^4$ for the ratio of the leading terms in $\eta_{A,B}$.

Finally, we have obtained  the contribution \eqref{eq:ffbremtot} of the bremsstrahlung mechanism which appears due to the account of the finite nuclear mass.
It turns out  that this contribution starts to dominate very soon when $\beta$ decreases.
This severely restricts the region of applicability of the results \eqref{eq:ffSpectrumAndCrossSection} and \eqref{eq:bfSpectrumAndCrossSection}.

\section*{Acknowledgement}
This work has been supported by Russian Science Foundation (Project No. 14-50-00080).
It has been also supported in part by RFBR (Grant No. 16-02-00103).


\end{document}